\documentclass[twocolumn,floatfix, showpacs]{revtex4}

\usepackage[final]{epsfig}
\usepackage{amsmath}

\begin{document}
 
\title{Energy dependence of the dijet imbalance in Pb-Pb collisions at 2.76 ATeV}
 
\author{Thorsten Renk}
\email{thorsten.i.renk@jyu.fi}
\affiliation{Department of Physics, P.O. Box 35, FI-40014 University of Jyv\"askyl\"a, Finland}
\affiliation{Helsinki Institute of Physics, P.O. Box 64, FI-00014 University of Helsinki, Finland}

\pacs{25.75.-q,25.75.Gz}

\begin{abstract}
The appearance of monojets is among the most striking signature of jet quenching in the context of ultrarelativistic heavy-ion collisions at the Large Hadron Collider. Experimentally, the asymmetry of back-to-back jets is quantified in terms of the dijet imbalance distribution by the ATLAS and CMS collaborations. Recently, the CMS collaboration has also studied the trigger jet momentum ($P_T$) dependence of the imbalance in the range between 120 and 500 GeV which is found to decrease with jet $P_T$. In this work, results from the in-medium shower code YaJEM are compared with this data set. These results suggest that the main effects observed in the data are the kinematical collimation of jets and the increase in the probability to produce more collimated quark jets with jet $P_T$, whereas there is no indication for any non-trivial energy dependence of the shower-medium interaction mechanism itself. The data furthermore can rule out models in which the jet shape is collimated due to the medium modification.

\end{abstract}
 
\maketitle

\section{Introduction}

Hard partonic reactions in perturbative Quantum Chromodynamics (pQCD) lead (to leading order in the strong coupling $\alpha_s$) to highly virtual back-to-back parton configurations which subsequently evolve as parton showers, hadronize and are then experimentally observed as back-to-back jet events. If such a  hard process takes place in a medium as created in ultrarelativistic heavy-ion collisions, the evolving shower interacts with the medium and thus the properties of the jet are altered.

Dijet imbalance observations as have been made by the ATLAS \cite{ATLAS} and CMS \cite{CMS} collaborations at the Large Hadron Collider (LHC) for 2.76 ATeV Pb-Pb collisions are an attempt to quantify the medium modification of the shower and several theoretical computations have been confronted with this data so far (see e.g. \cite{Dijets-Qin,Dijets-Vitev,Dijets-Martini}). Since energy-momentum conservation guarantees that the shower-initiating partons are balanced in momentum (up to intrinsic $k_T$ generated in the initial state partonic wave function and hard gluon radiation leading to three-jet topologies), the momentum imbalance is chiefly a statement about how much of the original parton momentum is reconstructed given a particular jet definition. This in turn means that event-by-event fluctuations of the observed energy-momentum flow in a jet determine the imbalance distribution in a crucial way.

Such fluctuations can have different sources: 1) fluctuation in the detector response 2) dynamical fluctuations of the parton shower pattern in vacuum 3) fluctuations of the soft background clustered into the jet in the presence of a medium and 4) modifications of the shower evolution by the medium. 1) and 2) lead to sizeable dijet imbalances even in p-p collisions, and 3) increases this imbalance to some degree even in the absence of final state shower-medium interactions. Any computation of the final state interaction requires hence a reasonable account of the fluctuations in the baseline 1) - 3) , as background fluctuations are potentially capable of masking the desired signal \cite{BgFluct}. The presence of background fluctuation is part of the reason for the lack of tomographic sensitivity of the set of first imbalance measurements as discussed in \cite{Dijets-Renk}.

The CMS collaboration has subsequently also measured the $P_T$ dependence of of the imbalance distribution. This data has the potential to overcome the lack of tomographic sensitivity mentioned above. The reason is that the different sources of fluctuations are expected to have a different $P_T$ dependence --- while detector response parametrically scales $\Delta P_T^{det} \sim \sqrt{P_T}$, the background medium fluctuations occur at a constant scale $\Delta P_T^{bg}$. The dynamical fluctuations due to the vacuum radiation have no simple form and depend on the jet definition. The main physics effect is the increased transverse collimation of jets at higher $P_T$ both due to the kinematic forward boost and due to the increased probability of producing (narrower) quark jets, thus for given jet cone radius the fraction of jet momentum reconstructed inside the cone is expected to increase with $P_T$. Finally, the $P_T$ dependence of the medium modification depends on the precise scenario assumed for shower-medium interaction and is thus in principle tested by the data.

The aim of this work is to illustrate these points using the example of the in-medium shower evolution Monte-Carlo (MC) code YaJEM \cite{YaJEM1,YaJEM2}. This code can be run simulating different physics scenarios, here we compare a multi-observable constrained scenario dominated by radiative energy loss with a small component of elastic energy transfer into the medium (YaJEM-DE) \cite{JetQuenchingPhysics} where jet shape is widened due to the interaction with the medium with a pure elastic drag scenario (YaJEM-E) \cite{YaJEM2} in which the medium modification collimates the jet shape which leads to a different prediction for medium-induced out of cone fluctuations. After a short overview over the model framework, results for the energy dependence of the dijet imbalance are presented and compared with the CMS data. This is followed by a discussion of the findings.

\section{The model}

The dijet imbalance is computed event by event using a MC strategy. The framework is described in detail in \cite{Dijets-Renk} where it is applied to the ATLAS dijet imbalance measurement, here we summarize the essential points and highlight the physics relevant to understand the results. The basic outline of the modelling procedure is as follows: We first generate a hard back-to-back parton configuration based on a leading order (LO) pQCD code \cite{Dihadrons}, sampling both parton type and momentum. To effectively account for initial state effects and next-to-leading order effects, an intrinsic ${\bf k_T}$ imbalance with a Gaussian distribution is introduced on the parton level (i.e. partons are never exactly back-to-back). The partonic configuration is, distributed with the transverse binary collision profile, embdedded into a hydrodynamical medium which is constrained by reproducing LHC bulk multiplicity and spectra \cite{LHCspectra} (note however that the results of \cite{Dijets-Renk} strongly suggest that the dijet imbalance is insensitive to even gross features of the medium model beyond the mean density). 

\begin{figure*}[htb]
\begin{center}
\epsfig{file=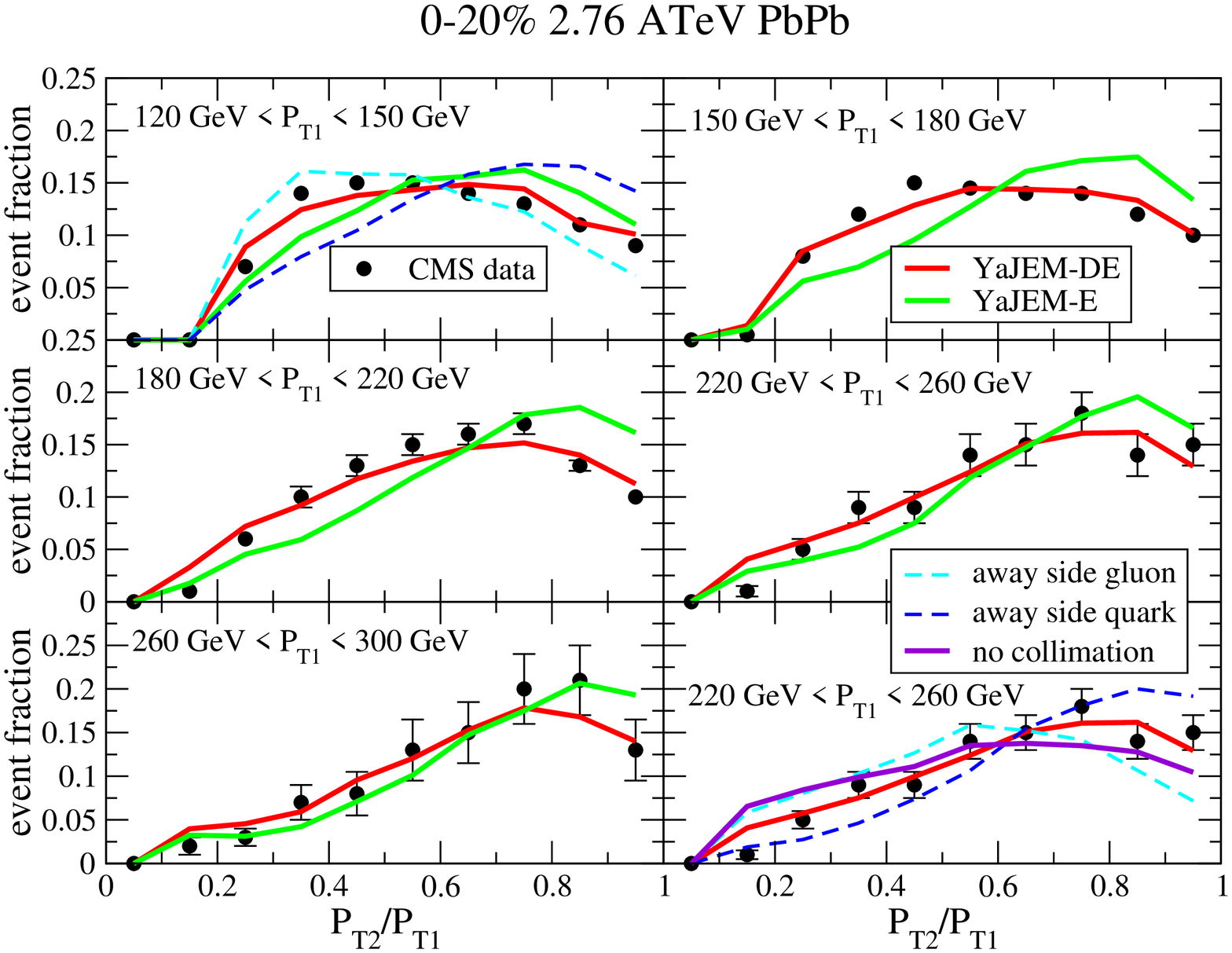, width = 14cm}
\end{center}
\caption{\label{F-1}Dijet momentum ratio distribution as measured by CMS and as computed both in YaJEM-DE and YaJEM-E for different trigger jet momentum ranges in central 2.76 ATeV Pb-Pb collisions. Shown in two panels is also the away side parton channel decomposition and the effect of kinematical collimation (see text), all compared with CMS data \cite{Dijets-Edep} (note that jet definitions are not identical).}
\end{figure*}

Based on the parton path through the evolving medium, we next utilize the code YaJEM \cite{YaJEM1,YaJEM2} to compute the in-medium parton shower. The parton-medium interaction in YaJEM is cast into the form of transport coefficients at the parton position $\zeta$ which are $\hat{q}(\zeta)$ (mean virtuality transfer to a parton per unit pathlength) and $\hat{e}(\zeta) $ (mean energy transfer from parton to the medium per unit pathlength). These are assumed to relate to the hydrodynamical energy density $\epsilon$ as

\begin{equation}
\label{E-qhat}
\hat{q}[\hat{e}](\zeta) = K_R[K_E] \cdot 2 \cdot \epsilon(\zeta)^{3/4} (\cosh \rho(\zeta) - \sinh \rho(\zeta) \cos\psi).
\end{equation}

where $\rho$ is the transverse flow rapidity of matter at position $\zeta$ and $\psi$ is the angle between flow and parton direction. $K$ and $K_E$ are adjustible parameters determining the overall strength of the interaction between partons and the medium. Note that YaJEM utilizes these transport coefficients to explicitly change parton kinenatics in the evolving shower in contrast to other in-medium shower models \cite{BW,Q-PYTHIA} which change the probability distributions for soft gluon emission rather than kinematics.

The ratio of $K_R$ and $K_E$ determines the relative contributions of medium-induced radiation and elastic energy loss into the medium. Here, we explore two scenarios:
The default scenario, YaJEM-DE utilizes $K_E = 0.1 K, K_R = 0.9 K$, i.e. about a 10\% contribution of elastic energy loss. This scenario is the result of a combined analysis of multiple RHIC and LHC single hadron and hard dihadron correlation observables \cite{JetQuenchingPhysics,YaJEM-DE}. It leads to a substantial broadening of jet shapes below $\sim 4$ GeV (i.e. at a scale comparable with the medium temperature) and leaves the jet structure almost unmodified at higher momenta (cf. the scenario RAD in \cite{JetShapes}).
We contrast this with a pure elastic drag model where $K_E = K', K_R = 0$ in which the medium does not induce any radiation but in which a substantial momentum flow exists from the shower into the medium. This scenario leads to a collimation of jet shapes (cf. the scenario DRAG in \cite{JetShapes}). Note that this scenario can be ruled out by several other high $P_T$ observables \cite{JetQuenchingPhysics}, however we use it to illustrate it to show the sensitivity of the energy dependence of the dijet imbalance to the jet shape.

$K$ (and $K'$) are fixed as follows: $K$ is determined by the requirement that $R_{AA}$ of single inclusive charged hadrons in central 2.76 ATeV PbPb collisions agrees with the data around 60 GeV (about the kinematical region where the lowest energy jet measurements are done). This leads to a $R_{AA}^{jets} = 0.48 \pm 0.02$ in line with ATLAS measurements of $R_{CP}$ \cite{Dijets-Renk} and is consistent with $K$ determinations from RHIC data within the uncertainty of extrapolating a hydrodynamical model from RHIC to LHC \cite{LHCspectra}. Since YaJEM-E is not consistent with other observables, $K'$ is selected to give the same trigger rate (i.e. the same $R_{AA}^{jets}$) as YaJEM-DE.

In order to compare with the measurement, the partonic output of the MC code in terms of event records must then be converted into jets. Hadronization is done using the Lund model \cite{Lund}. Since current jet measurements \cite{ATLAS,CMS} are not corrected for effects like detector response fluctuations or background fluctuations, a rigorous comparison of theory with the data would require the following steps: 1) generation of a bulk event using for instance hydrodynamics with event by event fluctuating initial conditions 2) generation of a hard event inside the same bulk event (since there is an observable correlation between medium fluctuation hotspots and hard vertices \cite{Fluctuations}, the hard and soft physics can not be modelled independently in a rigorous treatment) and evolution of the perturbative shower inside the medium 3) computation of the back-reaction of the shower to the medium to account for correlated soft physics inside the jet area (for instance elastic recoil of medium constituents) 4) simulating the detector response to turn particle event record information into calorimeter tower information 5) clustering and analysis of the calorimeter-level (or particle flow level) event as in the experimental procedure.

So far, no theoretical model has realized this full, numerically very involved workflow (note also that detector response simulation codes are usually not available outside the experimental collaborations). Nevertheless, it is important to construct a meaningful baseline calculation before final state effects are applied which at least contains qualitatively the phenomena affecting the experimental measurement and quantitatively reproduces strength and distribution of the fluctuations. A comparison with the published baseline (here PYTHIA + HYDJET in \cite{Dijets-Edep}) suggests that fluctuations are significantly more complex than Gaussian, thus this is a non-trivial task. For this investigation, a baseline which reproduces the fluctuations in the published experimental baseline is constructed as follows: A hadron contributes to the jet momentum if:  a) the hadron is at an angle of $R=0.3$ or less with the jet axis b) the hadron species is one of $\pi^+, \pi^-, \pi^0, K^+, K^-, p, \overline{p}, \gamma$ c) hadron $P_T > 1$ GeV. The resulting jet momentum gets a contribution from random background fluctuations of a Gaussian distribution centered at zero with 14 GeV width (note that this is different from what has been used for ATLAS $R=0.4$ jets in \cite{Dijets-Renk} --- since the jet definitions of ATLAS and CMS are not the same, there is no reason for the numbers to agree). Following the procedure applied by CMS, all jets with an energy below 30 GeV are discarded.

While these conditions are certainly not identical to the result of a PYTHIA + HYDJET event propagated through a GEANT detector simulation and clustered by the anti-$k_T$ algorithm \cite{antiKT}, the important thing to note is that they reproduce the pattern of fluctuations in the reconstruncted jet $P_T$ baseline well. Thus, while no strictly rigorous comparison with data is possible, one has a reasonable baseline to study trends and scaling patterns of the medium modifications.

To sumarize, fluctuations of the reconstructed jet energy within the model are created by the following mechanisms: 1) intrinsic ${\bf k_T}$ imbalance of the partonic initial state 2) dynamical fluctuations by vacuum radiation leading to e.g. hadron production  outside the jet cone 3) uncorrelated medium background energy fluctuations imposed in a Gaussian approximation  4) medium induced additional broadening of the jet shape, leading to enhanced out of cone and soft hadron production. Note that only the last point is characteristic for the shower-medium interaction model.

\section{Results}

The main result of this work is shown in Fig.~\ref{F-1} where the momentum ratio distribution of associate jet momentum $P_{T2}$ and trigger jet momentum $P_{T1}$ is shown for various trigger momentum ranges between 120 and 260 GeV. The general trend in both model calculations and the data is that jets become less imbalanced at higher $P_T$. While the constrained scenario YaJEM-DE is in reasonable agreement with the CMS data throughout the entire momentum range investigated here, the jet collimating scenario YaJEM-E tends to underpredict the asymmetry seen in the data in the lower trigger momentum ranges. For higher trigger momenta the quality of the description of the data becomes comparable in both scenarios. 

At first glance, this seems to contradict the findings of \cite{Dijets-Renk} where no distinction between YaJEM-DE and YaJEM could be established. Note however the difference in cone radius $R=0.4$ vs. $R=0.3$ as used here --- since the two scenarios differ in jet shape, it is expected that they do not respond in the same way to a change in cone radius.

The non-trivial trigger $P_{T1}$ dependence of the two scenarios supports the expectation that this set of data probes the momentum dependence of various fluctuation-generating mechanisms in a non-trivial way: Just based on the 260 -- 300 GeV result, it would be difficult to distinguish the two scenarios, however the full momentum dependence provides the information to overcome this problem.

To further investigate the physics underlying the results, for two trigger ranges results with the away side shower-initiating parton forced to a gluon or to a quark computed in YaJEM-DE are shown. These results are in line with the expectation that the broader transverse shape of gluon jets should lead to a stronger asymmetry by inducing more out of cone radiation. While at low trigger $P_{T1}$ the full result is somewhat closer to the gluon result, at high $P_{T1}$ this is no longer true and the full curve is about the average between quark and gluon result. This confirms the expectation that the rise in the relative fraction of quark jets with jet momentum contributes to the collimation of observed jets and hence to a reduction in the observed imbalance. 

A different effect is the kinematical collimation of high $P_T$ jets due to the forward boost of evolving shower. We can assess the role of this effect by forcing jets triggered at 220 GeV to have the same jet shape distribution as jets at 120 GeV. If this is done, as expected an increased imbalance is found which is of the same order as what is induced by the gluon to quark transition. We can conclude from this that the decrease of the dijet imbalance with trigger momentum is caused equally by both effects.

\section{Conclusions}

The results suggest that the observed change in the dijet imbalance as a function of trigger jet momentum can be accounted for in a scenario that is well-constrained by other observables and can be explained in terms of essentially known physics, i.e. the increased collimation of jets due to kinematics and a transition to a less gluon-dominated regime. In other words, no particular special or exotic assumptions beyond kinematics need to be made for the $P_T$ dependence of the shower modification by the medium itself.

The failure of YaJEM-E to account for the data in part of the momentum range indicates that the jet $P_T$ dependence of the imbalance is as expected sensitive to the jet shape and its distortion by the medium modification. Comparison with \cite{Dijets-Renk} suggests that smaller cone sizes may offer observables more sensitive to the transverse structure generated by shower-medium interaction scenarios.

In summary, the reasonable agreement between the trends of the imbalance evolution as seen in the data and in YaJEM-DE strengthen the case for the interpretation of the combined data of high $P_T$ observables as being dominated by pQCD radiative energy loss with a small fraction of elastic energy transfer into the medium.

\begin{acknowledgments}
 
This work was supported by the Academy Researcher program of the Finnish Academy (Project 130472). Discussions with  Matteo Cacciari Christof Roland and Gunther Roland are gratefully acknowledged. Part of the numerical computations was carried out with generous support by Helen Caines on the {\bf bulldogk} cluster at Yale University.
 
\end{acknowledgments}


\begin{thebibliography}{99}

\bibitem{ATLAS}
 G.~Aad {\it et al.} [ Atlas Collaboration ],
  Phys.\ Rev.\ Lett.\  {\bf 105 } (2010)  252303.

\bibitem{CMS}
 S.~Chatrchyan {\it et al.} [ CMS Collaboration ],
  Phys.\  Rev.\  {\bf C84 } (2011)  024906.

\bibitem{Dijets-Qin}
G.~-Y.~Qin and B.~Muller,
  Phys.\ Rev.\ Lett.\ \ {\bf 106} (2011) 162302.

\bibitem{Dijets-Vitev}
 Y.~He, I.~Vitev and B.~-W.~Zhang,
 1105.2566 [hep-ph].

\bibitem{Dijets-Martini}
  C.~Young, B.~Schenke, S.~Jeon and C.~Gale,
  1103.5769 [nucl-th].

\bibitem{BgFluct}
 M.~Cacciari, G.~P.~Salam and G.~Soyez,
  Eur.\ Phys.\ J.\ C {\bf 71} (2011) 1692

\bibitem{Dijets-Renk}
  T.~Renk,
  1202.4579 [hep-ph].

\bibitem{Dijets-Edep}
 S.~Chatrchyan {\it et al.}  [CMS Collaboration],
  1202.5022 [nucl-ex].

\bibitem{JetQuenchingPhysics}
  T.~Renk,
  1112.2503 [hep-ph].

\bibitem{YaJEM1}
  T.~Renk,
  Phys.\ Rev.\  C {\bf 78} (2008) 034908.

\bibitem{YaJEM2}
 T.~Renk,
  Phys.\ Rev.\  C {\bf 79} (2009) 054906.

\bibitem{Dihadrons}
T.~Renk and K.~Eskola,
  Phys.\ Rev.\ C {\bf 75} (2007) 054910.

\bibitem{LHCspectra}
  T.~Renk, H.~Holopainen, R.~Paatelainen and K.~J.~Eskola,
  Phys.\ Rev.\ C {\bf 84} (2011) 014906.

\bibitem{BW}
 N.~Borghini and U.~A.~Wiedemann,
  hep-ph/0506218.

\bibitem{Q-PYTHIA}
 N.~Armesto, L.~Cunqueiro and C.~A.~Salgado,
  Eur.\ Phys.\ J.\  C {\bf 63} (2009) 679.

\bibitem{YaJEM-DE}
 T.~Renk,
  Phys.\ Rev.\ C {\bf 84} (2011) 067902.

\bibitem{JetShapes}
 T.~Renk,
  Phys.\ Rev.\ C {\bf 80} (2009) 044904.

\bibitem{Fluctuations}
  T.~Renk, H.~Holopainen, J.~Auvinen and K.~J.~Eskola,
  1105.2647 [hep-ph].

\bibitem{Lund}
  B.~Andersson, G.~Gustafson, G.~Ingelman and T.~Sjostrand, Phys. Rep. {\bf 97}
(1983) 31.

\bibitem{antiKT}
  M.~Cacciari, G.~P.~Salam and G.~Soyez,
  JHEP {\bf 0804} (2008) 063.

\end{thebibliography}
\end{document}